\documentclass[aps,prl,epsfig,floats,twocolumn,superscriptaddress,amssymb,amsmath,floatfix,showpacs]{revtex4}
\usepackage{graphicx}
\usepackage{color}
\begin{document}
\title{Soret Motion of a Charged Spherical Colloid}

\author{Seyyed Nader Rasuli}
\email{rasuli@mail.ipm.ir}\affiliation{School of the Physics, IPM
(Institute for Theoretical Physics and Mathematics), P.O.Box
19395-5531, Tehran, Iran} \affiliation{Institute for Advanced Studies in Basic Sciences, Zanjan 45195-1159,
Iran}

\author{Ramin Golestanian}
\email{r.golestanian@sheffield.ac.uk} \affiliation{Department of
Physics and Astronomy, University of Sheffield, Sheffield S3 7RH,
UK}

\date{\today}

\begin{abstract}
The thermophoretic motion of a charged spherical colloidal particle
and its accompanying cloud of counterions and co-ions in a
temperature gradient is studied theoretically. Using the
Debye-H\"uckel approximation, the Soret drift velocity of a weakly
charged colloid is calculated analytically. For highly charged
colloids, the nonlinear system of electrokinetic equations is solved
numerically, and the effects of high surface potential,
dielectrophoresis, and convection are examined. Our results are in
good agreement with some of the recent experiments on highly charged
colloids without using adjustable parameters.
\end{abstract}


\maketitle

\paragraph{Introduction.} A temperature gradient applied to a fluid
mixture causes relative transformation of its components; some
condensing in the hotter and some in the colder side. This
phenomenon, known as the Soret effect or thermophoresis, has been
studied for nearly 150 years \cite{Ludwig} and has been observed in
a variety of systems
\cite{Lenglet,Piazza-PRL,Iacopini,DB,Wigand,putnam}. While the
existence of such a {\em response} can be well formulated in the
framework of non-equilibrium thermodynamics \cite{de Groot}, in many
cases its microscopic nature remains poorly understood
\cite{Ruck,Morozov,general,Dhont,Astumian,Fayolle}. The
manifestation of this effect in charged colloidal solutions is
particularly puzzling: while it is usually observed that colloids
condense in the colder side, experiments with opposite results also
exist \cite{Lenglet} and it appears that the tendency could change
with the variation of salt concentration and temperature
\cite{Piazza-PRL,Iacopini}.

In a pioneering work in 1981, Ruckenstein suggested a model for
Soret motion of a single charged colloidal particle \cite{Ruck}.
Using Boltzmann distribution for ion densities and the
Debye-H\"uckel approximation for the electrostatic potential, he was
able to solve the hydrodynamics equations and find the drift
velocity of a colloid in a temperature gradient.
The 2002 experiment by Piazza and Guarino on Sodium Dodecyl Sulfate
(SDS) micelles solutions provided an opportunity for verification of
this model \cite{Piazza-PRL}. The SDS micelles in the experiment
were highly charged, as they had an estimated saturated structural
charge of $Z \simeq 75$ (in electronic unit) \cite{SDS structural
charge} and a radius of $a=2.5$ nm, and the thickness of their
double-layer---as set by the Debye length---varied between $0.4$ nm
to $2.5$ nm \cite{Piazza-PRL}. The experiment was therefore outside
the region of validity of the Ruckenstein's theory, which was
restricted to weakly charged colloids with thin double-layers.
Piazza and Guarino showed that it was possible to get a reasonable
fit to the experimental data for the dependence of the Soret
coefficient on the salt concentration using Ruckenstein's formula,
provided they assumed an increased radius of $a=3.5$ nm and a
reduced charge of $Z=17$ \cite{Piazza-PRL}. The apparent smaller
charge of the colloid could be interpreted as the renormalized
charge as described by Alexander {\em et al.} \cite{Alexander}.
However, this attribution is not entirely justified as the concept
of effective charge is defined through the asymptotic form of the
electrostatic potential of the colloid far away from its surface
\cite{Alexander}, while Ruckenstein's model deals with the electric
field inside the thin double-layer. In fact, the value of the
effective charge $Z=17$ used by Piazza and Guarino are obtained from
measurements on inter-colloidal interactions \cite{SDS structural
charge}, and electrophoretic observations on SDS micelles does not
seem to confirm this value \cite{Stiger}. More recently, Putnam {\em
et al.} have performed experiments on highly charged T4 lysozymes
and found results that do not seem to be explained satisfactorily by
any of the existing theories on single-colloid thermophoresis
\cite{putnam}. Another recent experiment by Duhr and Braun \cite{DB}
probed charged particles with low surface potentials, and found a
new scaling relation for Soret motion of charged colloid which does
not agree with Ruckenstein's prediction. They showed that it is
possible to explain their observations, using the Gibbs enthalpy of
the charged colloid \cite{DB} (or the irreversible work needed to
construct it \cite{Dhont}). The controversy was later clarified to
some extent but the work of Astumian \cite{Astumian}, who suggested
that while Ruckenstein's model deals with the deterministic steady
motion of a charged colloid, the theory by Duhr and Braun addresses
its fluctuation--induced stochastic motion, which is a separate
contribution, and a complete picture should involve both of these
aspects simultaneously.

Here, we focus on the deterministic motion of a charged spherical
particle in a temperature gradient, and consider both weakly charged
and highly charged cases taking into account effects such as the
temperature dependence of solvent electric permittivity, convection,
and nonequilibrium co/counterion redistribution. We examine the
behavior of system using Debye-H\"uckel approximation for weakly
charged colloids, and provide an analytical result for colloid drift
velocity with {\em arbitrary} double-layer thicknesses. For highly
charged colloids, we solve the nonlinear set of coupled equations
numerically, and find that the Soret coefficient has a non-monotonic
dependence on the surface (zeta) potential of the colloids, an
effect that has been indeed observed by Putnam {\em et al.}
\cite{putnam}. We also examine the relative importance of various
contributions in the different regimes. For the experiment with SDS
micelles \cite{Piazza-PRL}, we show that our systematic approach can
yield results in reasonable agreement with the experimental data,
using a realistic bare value for the micellar charge.

\paragraph{The model.} We consider an ionic solution of different
species with concentration $C_i$ and valence $q_i$, which create an
electrostatic potential $\phi$ through Poisson's equation
\begin{math}
-\vec{\nabla} \cdot \varepsilon \vec{\nabla} \phi =4 \pi \sum_i q_i
C_i
\end{math}.
This solution is subject to a gradient in temperature which causes
$\varepsilon$ to change in space. We consider weakly varying
temperature fields, so use linear response theory to study the
deformation of the double-layer and solvent flow. The Soret motion
of a charged colloid is a steady motion which means that in the
colloid framework, we should look for a stationary solution for the
fluid velocity field $\vec{V}$. This fluid velocity is governed by
the Navier-Stokes equation and the incompressibility constraint
$\vec{\nabla} \cdot \vec{V}=0$. In the limit of low Reynolds number,
this yields the stationary Stokes equation:
\begin{math}
-\eta \nabla^2 \vec{V}=-\vec{\nabla} P- \vec{\nabla}\phi \sum_i q_i
C_i+\vec{f}
\end{math}
Here $\eta$ is the viscosity of the solvent, and $P$ is its
pressure. The body force $\vec{f}$ is the dielectrophoretic force,
which comes from the net force experienced by electric dipoles of
water molecules because of spatial variation in the electric field
\cite{epaps}. It reads \cite{Dielectrophoretic force}:
\begin{math}
\vec{f}=\vec{\nabla}\left(\frac{\varepsilon-1}{8
\pi}|\vec{E}|^2\right)-\frac{1}{8 \pi}|\vec{E}|^2
\vec{\nabla}\varepsilon
\end{math}
where $\vec{E}=-\vec{\nabla}\phi$. The first term on this expression
is a complete derivative and can be absorbed in the fluid pressure
({\em{i.e.}} $P \rightarrow P - \frac{\varepsilon-1}{8
\pi}|\vec{E}|^2$). Assuming that the solvent is incompressible ({\em
i.e.} $\rho = \rho_{0}$), changes in its permittivity
$\varepsilon(\rho_{0}, T)$ is due to the temperature gradient only:
\begin{math}
-\frac{1}{8
\pi}|\vec{E}|^2\vec{\nabla}\varepsilon=\alpha\frac{\varepsilon}{8
\pi}|\vec{E}|^{2} \frac{\vec{\nabla T}}{T}
\end{math}
where $\alpha=-\partial \ln \varepsilon/\partial \ln T$ is about
$1.35$ for water in room temperature \cite{CRC}. Finally, each ion
species is subject to a conservation law $\vec{\nabla} \cdot
\vec{J_i}=0$, with ionic current density:
\begin{equation}
\vec{J}_i=- D_i \vec{\nabla}C_i-\mu_i C_i q_i \vec{\nabla}\phi +
\vec{V} C_i - D_{i} C_{i} S_{T}^{\rm
ion}\vec{\nabla}T\label{eq:general Fick}
\end{equation}
where $\mu_i$ is the $i$th type of ions mobility and $D_i= \mu_{i}
k_{B} T$ is its diffusivity. $S_{T}^{\rm ion}$ is the
co/counter-ions Soret coefficient. Here we only focus on $1:1$
electrolytes, so we assume that coions and counterions have equal
Soret coefficients \cite{Cald}.

In the absence of any temperature gradient and fluid motion, Eq.
(\ref{eq:general Fick}) is simplified to its first two terms. Then,
$C_{i}=C_{0}\exp[-q_{i} \phi/k_{B}T]$ yields $\vec{J}_{i}=0$, and
satisfies $\vec{\nabla} \cdot \vec{J}_{i}=0$. In the presence of a
temperature gradient, however, not only we have to consider all of
Eq. (\ref{eq:general Fick}) terms, but also note that a Boltzmann
weight form for $C_{i}$ no longer makes $-D_{i}\vec{\nabla}C_{i}
-\mu_i C_i q_i \vec{\nabla}\phi$ vanish; instead, it yields $- D_i
C_{i} (q_{i} \phi/k_{B}T^2)\vec{\nabla}T$. We suggest to use the
following form
\begin{math}
C_i=C_0 \exp\left[-\frac{q_i \phi}{k_{\rm B} T}-(T-T_0)S_{T}^{\rm
ion}+ \left(\frac{T - T_{0}}{T_0}\right)\Omega_i \right],
\end{math}
where $\Omega_i$ measures the deviation of the concentration from a
Boltzmann weight form, and it contains contribution from the
convective term in Eq. (\ref{eq:general Fick}) as well as the
aforementioned term of $- D_i C_{i} (q_{i}
\phi/k_{B}T^2)\vec{\nabla}T$.

Our aim is to find the colloid drift velocity, so we focus on
the Stokes equation that governs fluid velocity. The electric and
dielectrophoretic forces on the right hand side are acting as source
terms that induce fluid motion. To first order in temperature
changes, the source terms simplify to
\begin{equation}
- \vec{\nabla}\phi_0 \sum_i q_i C_i^{0} \left[ \frac{q_i
\phi_0}{k_{B} T_0} - T_{0} S_{T}^{ion} + \Omega_i
\right]\frac{\delta T}{T_0} + \alpha\frac{\varepsilon
|\vec{E}_{0}|^{2}}{8 \pi} \frac{\vec{\nabla}T}{T_0},
\label{eq:governing Eq}
\end{equation}
where $C_i^{0}=C_{0}\exp[-q_{i} \phi_{0}/k_{B}T_{0}]$ and
$\vec{E}_{0}$ are the non-perturbed values of ion density and
electric field. Here, we have also extracted another complete
derivative ({\em i.e.}$-\vec{\nabla}\sum_i q_i C_i^0 \delta\phi$),
which can be absorbed in the fluid pressure term \cite{epaps}.

\paragraph{Weakly Charged Colloids.} For $q \phi/k_{\rm B}T \ll 1$,
we can use the Debye-H\"uckel approximation and solve this system of
equations analytically \cite{epaps}. The drift velocity of a colloid
with radius $a$ is found as
\begin{eqnarray}
\vec{V}_{\rm drift}&=&\frac{-\varepsilon
\phi_{S}^2}{48\pi}\frac{\vec{\nabla}T}{\eta T_{0}}
 \label{Particle drift DH} \\
 &\times&\left\{(1+T_{0}S_{T}^{\rm ion}) F(\kappa a)-G(\kappa
a)+\alpha\left[2-F(\kappa a)\right] \right\},\nonumber
\end{eqnarray}
where $\phi_{S}=Z q/\varepsilon a (1+ \kappa a)$ is the zeta
potential of the surface of the colloid and $\kappa=\sqrt{8 \pi q^2
C_0/\varepsilon k_{\rm B} T_0}$ is the inverse Debye length
\cite{colloid}. In this equation, $F(x)=2 x-4 x^2 e^{2 x} E_1(2 x)$
and $G(x)=\frac{x}{6} [x (1+x)(12-x^2) e^x E_1(x)+8-11 x+x^3-24 x
e^{2 x} E_1(2 x)]$, with $E_1(x)=\int_{x}^{\infty} e^{-s} ds/s$.
Equation (\ref{Particle drift DH}) is arranged in the form of
Ruckenstein's result for colloid drift velocity \cite{Ruck}
multiplied by a correcting factor. Each term in this correcting
factor corresponds to one of the source terms in Eq.
(\ref{eq:governing Eq}) ({\em i.e.} $\alpha [2-F(\kappa a)]$
corresponds to the dielectrophoretic term, $-G(\kappa a)$ to
$\Omega_{i}{\delta T}/{T_0}$ term, and $T_0 S_{T}^{\rm ion}F(\kappa
a)$ to ions Soret term). The remaining $F(\kappa a)$ corresponds to
the $- \vec{\nabla}\phi_0 \sum_i q_i C_i^{0} \left[ \frac{q_i
\phi_0}{k_{B} T_0} \right]{\delta T}/{T_0}$) term, which corrects
Ruckenstein's result \cite{Ruck} for arbitrary double-layer
thicknesses.

The limit of a thick double-layer ({\em i.e.} $\kappa a \ll 1$) or
low ionic strength is particularly interesting, as for $\kappa a=0$
we have $F(\kappa a)=G(\kappa a)=0$ \cite{epaps}, and the only
contribution in the correcting factor of Eq. (\ref{Particle drift
DH}) will be $2 \alpha$. It means that with low ionic density, the
dielectrophoretic force that is the force on water molecules
\cite{epaps}, plays the dominant role in the phenomenon. In
addition, Ruckenstein's formula will be still applicable, if we
multiply it with a constant $2 \alpha$, which can be presented in
terms of a renormalized surface potential $\phi_{S}^{'}=\sqrt{2
\alpha} \phi_{S}$ or charge $Z^{'}=\sqrt{2 \alpha} Z$
\cite{footnote}.

\begin{figure}
\includegraphics[width=0.90\columnwidth]{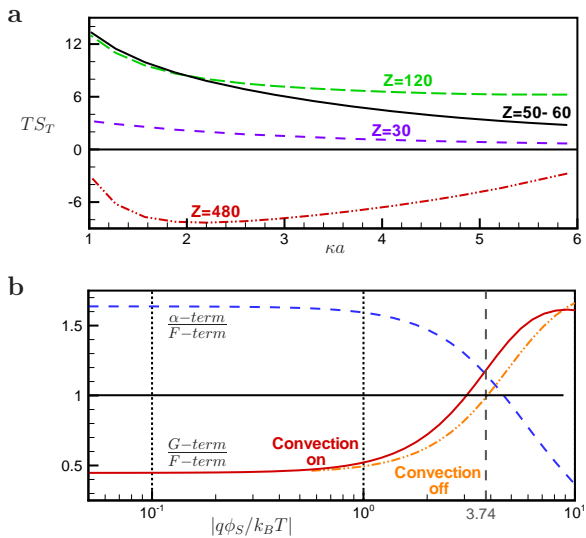} \caption{
(color online). (a) $T S_{T}$ versus $\kappa a$ for a colloid with
radius $a=2.5$nm, but different charges. Mean temperature is
$T=25^\circ$C, and we extrapolated existing data \cite{Cald} to
obtain (Na$^{+}$ and CL${^-}$) ions Soret coefficient: $S_{T}^{\rm
ion}=0.99-2\times10^{-3}$ K$^{-1}$ (depending on salt density). (b)
Ratios of the dielectrophoretic term ($\alpha$-term) to the $F$-term
(dashed blue line) and $G$-term/$F$-term (solid red line), versus
$|q\phi_{S}/k_{B}T|$, for $\kappa a=2$. $|q\phi_{S}/k_{B}T|=3.74$
corresponds to the $Z=50-60$, our suggested charge for a micelle in
the experiment of Ref. {\cite{Piazza-PRL}}. The dashed-dotted
(orange) curve corresponds to when the convection term is
artificially switched off.}\label{fig:potential dependence}
\end{figure}

\paragraph{Highly Charged Colloids.} The Debye-H\"uckel approximation is not
valid for highly charged colloids and one needs to fully take
account of the nonlinearity of the electrostatics. To this end, we
have solved the above governing equations numerically. Figure
\ref{fig:potential dependence}a shows the Soret coefficient $T S_T
\sim V_{\rm drift}$ as a function of $\kappa a$ for various values
of structural charge, corresponding to the SDS micelles used in the
experiment of Ref. \cite{Piazza-PRL}. We find that while the Soret
coefficient initially increases with the structural charge---in
agreement with the Debye-H\"uckel limit behavior where it is
proportional to $\phi_{S}^2 \sim Z^{2}/(1 + \kappa a)^2$---this
trend is reversed at sufficiently high values of the charge and the
Soret coefficient starts to decrease until it changes sign and
becomes negative. We can understand this behavior better, if we
track the contributions of the different source terms in Eq.
(\ref{eq:governing Eq}) to the Soret coefficient. Following our
analytical approach [that led to Eq. (\ref{Particle drift DH})], we
name the contribution of $- \vec{\nabla}\phi_0 \sum_i q_i
C_i^{0}(q_i \phi_0/k_{B} T_0)(\delta T/T_0)$ term in Eq.
(\ref{eq:governing Eq}) as the $F$-term, that of the
dielectrophoretic term as the $\alpha$-term, and finally that of
$\Omega_i$'s as the $-G$-term (note minus sign of $G(\kappa a)$ in
Eq. (\ref{Particle drift DH})). Figure \ref{fig:potential
dependence}b shows the relative importance of these contributions as
a function the surface potential. While for weakly charged colloids
({\em i.e.} $|q \phi_{S}/k_{B}T| \ll 1$), the $\alpha$-term is the
dominant contribution, for $|q \phi_{S}/k_{B}T| \geq 1$, the
situation changes in favor of the $G$-term, which eventually leads
to negative Soret coefficient at high surface potentials.
Interestingly, for the surface potentials attributed to the SDS
micelles of Ref. \cite{Piazza-PRL}, the two terms balance each
other.

\begin{figure}
\includegraphics[width=0.80\columnwidth]{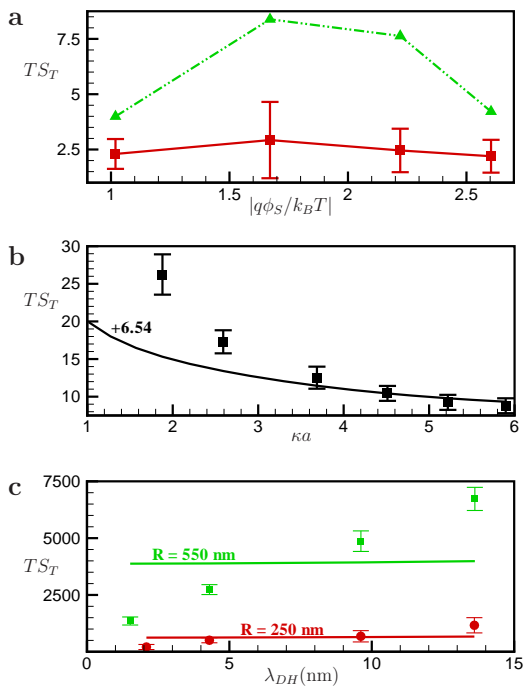}
\caption{(color online). (a) Comparison of Putnam {\em et al.}
\cite{putnam} data (red) for mutant T$4$L lysozymes with
$Z=+3,+5,+7$,$+9$, and $R_{h} \simeq 1.8$nm, with our numerical
prediction (dashed-dotted green curve). (b) Comparison of the
experimental data of Piazza and Guarino \cite{Piazza-PRL} with our
theoretical prediction. Depending on salt concentration, we assume a
varying micelle charge of $Z=50-60$ \cite{epaps}, a varying ions
Soret Coefficient $S_{T}^{\rm ion}=0.99-2\times10^{-3}$ K$^{-1}$,
and a fixed ionic diffusion coefficient $D= 1.33 \times 10^9 {\rm
nm}^2/{\rm s}$ \cite{our-PRE}. Our data is lifted by $+6.547$, to
obtain the best fit with four last points. (c) Comparison of our
analytic theory with Duhr and Braun data \cite{DB} for polystyrene
spheres with different radii.} \label{fig:ST-exp-th}
\end{figure}

A similar non-monotonic behavior has recently been observed by
Putnam {\em et al.} for lysozyme proteins with various surface
charges/potentials \cite{putnam}. Figure \ref{fig:ST-exp-th}a shows
a comparison between their observed $T S_{T}$ versus lysozyme
surface potential, and the prediction of our theory without any
adjustable parameters. In this calculation, we have ignored the
contribution of the Soret coefficient of the ions, as we were not
able to obtain relevant experimental values for these quantities.
The encouraging agreement suggests that for highly charged colloids
it will not be sufficient to treat the double layer close to the
surface within the equilibrium Poisson-Boltzmann formulation, and it
is necessary to fully take account of the nonlinear nonequilibrium
effects in the profile of the ion cloud around the colloid.

It is also interesting to probe the role of convection in our
results. The convection term in Eq. (\ref{eq:general Fick}) affects
$\Omega_{i}$'s and consequently the $G$-term. The dashed-dotted
(orange) curve in Fig. \ref{fig:potential dependence}b shows the
ratio $G$-term/$F$-term, with convection artificially turned off. We
observe that while for $|q \phi_{S}/k_{B}T| \ll 1$ the convective
term has a small effect (proportional to $|q \phi_{S}/k_{B}T|^2$
\cite{epaps}), for higher surface potentials the convective motion
will be non-negligible, but always finite.

We also do numerical calculation for experiment of Piazza and
Guarino, and use electrophoresis data \cite{Stiger} to find micelles
charge \cite{epaps}. The result is presented in Fig.
\ref{fig:ST-exp-th}b, which only shows an agreement with the
experiment for $\kappa a > 3$. Since we have established (see Fig.
\ref{fig:potential dependence}) that changing the structural charge
does not lead to a better harmony, we can conclude that effects
other than the ones considered here systematically should control
the behavior of the system for $\kappa a < 3$.

\paragraph{Discussion.}

The experiment performed by Duhr and Braun \cite{DB}, unlike most
other experiments, have been done for colloids within
Debye-H\"{u}ckel regime. Therefore, it is natural to expect that our
analytical results should agree with its findings. Figure
\ref{fig:ST-exp-th}c compares our result (based on Eq.
(\ref{Particle drift DH})) with its data, and does not show a
satisfactory agreement. We have examined possible corrections coming
from fluid slip velocity on the colloid surface and mismatch in the
thermal conductivities of the colloid and the fluid, and found that
they cannot improve the situation \cite{our-PRE}. To resolve this
discrepancy it has been suggested \cite{Astumian} that one should
differentiate between the regime where hydrodynamic or deterministic
components of the motion are dominant, and the regime in which the
system feels a local thermal equilibrium and it is the stochastic
motion which is the dominant player. It is expected that the
behavior of the Soret coefficient as a function of various
parameters such as the radius, temperature, etc. will be
significantly different in these two regimes. This means that a
theoretical formulation that can account for experiments
corresponding to the hydrodynamic or deterministic regime is almost
bound to fail to account for experiments in the stochastic regime as
they would be mutually exclusive, a view which is also supported by
recent experimental evidence \cite{Piazza new PRL}. Further work is
needed to fully clarify this picture, ideally in the form of a
theoretical formulation that includes both effects and can show the
crossover from one behavior to the other in a systematic way.

In conclusion, we have presented a systematic theoretical analysis
of the Soret motion of charged colloids both in the weakly and
strongly charged regimes. Using contributions from different
physical sources, we have found competing tendencies that are
entirely due to the nonlinearity of electrostatics and the
nonequilibrium redistribution of the ion clouds around the colloid.

We are grateful to E. Bringuier, R. Ejtehadi, F. Miri, and P. Pincus
for helpful discussions; and A. Naji, S. A. Putnam, and G. Wong for
their assistance.



\begin{thebibliography}{99}
\bibitem{footnote}
In some practical situations, we may use these renormalized surface
charge/potential for arbitrary $\kappa a$'s. It is because other
terms in the correcting factor of Eq. (\ref{Particle drift DH}) are
usually smaller than $2 \alpha$. For the experiment of Ref.
\cite{Piazza-PRL} for instance, $T S_{T}^{ion}$ varies between $0.3$
and $0.6$ (depending on salt density \cite{our-PRE}), then $(1 + T
S_{T}^{\rm ion} -\alpha)F(\kappa a)$ and $G(\kappa a)$ are small
enough that the deviation of the correcting bracket from $2 \alpha$
is less than $20\%$.
\end{thebibliography}
\end{document}